\documentclass[pra,aps,showpacs,twocolumn]{revtex4}
\usepackage{epsfig}

\begin{document}

\title{Generation of optical combs in a
whispering gallery mode resonator from a bichromatic pump}

\author{Dmitry V. Strekalov}
\author{Nan Yu}

\affiliation{Jet Propulsion Laboratory, California Institute of
Technology, 4800 Oak Grove Drive, Pasadena, California 91109-8099
}

\date{\today}

\begin{abstract}
An optical comb is shown to arise from a whispering gallery mode resonator pumped by two optical frequencies. Two externally excited modes couple due to Kerr nonlinearity to initially empty modes and give rise to new frequency components. This thresholdles process is much more efficient than the previously reported single-pump four-wave mixing. As a result, a few milliwatt pump is sufficient to generate strong secondary fields, that efficiently generate higher-order frequency components and so on, in a cascade process leading to an optical comb.
\end{abstract}


\maketitle

Optical frequency combs \cite{udem02,yebook} found their applications as optical clocks and frequency standards of very high accuracy \cite{Ma04}, including those for astronomy \cite{Li08Nature} and molecular spectroscopy \cite{thorpe06,diddams07}. They have also been proposed for use in quantum information processing \cite{zaidi08comb,menicucci08arX}. Four-wave mixing in a resonator with Kerr nonlinearity \cite{savchenkov04,kippenberg04} has been recently demonstrated \cite{del'haye07,savchenkov08} as one of the methods to generate optical combs. In this case the comb arises from direct or cascaded frequency conversion processes, which requires very high effective nonlinearities. The latter can be achieved by using the whispering gallery mode (WGM) resonators, known for their high $Q$ factors. 

\begin{figure}[htp]
\centerline{
\input epsf
\setlength{\epsfxsize}{3.8in} \epsffile{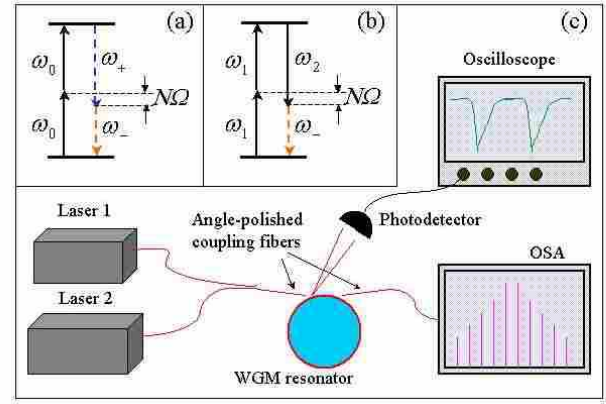} }
\vspace*{-0.7in}
\caption[]{\label{fig:setup} Energy diagrams of previously (a) and presently realized (b) four-wave mixing processes generating a frequency comb. The experimental setup (c).
}
\end{figure}

In monochromatically pumped Kerr-based optical comb sources two pump photons are converted into a quantum-correlated photon pair. This process, whose energy diagram is shown in Fig.~\ref{fig:setup}(a), populates the pair of modes adjacent to the pump as well as those two, three, etc., of the resonator free spectral ranges (FSR) away from the pump. To generate such a frequency comb in a WGM resonator, the pump has to exceed a certain power threshold $P_{th}$, inversely proportional to $Q^2$ \cite{Matsko05}. The factor $Q^2$ also determines the phase drift of the comb componets, via the Schawlow-Townes formula \cite{Matsko05}. On the other hand, the frequency span of the comb is limited by dispersion and therefore scales as the resonator linewidth, i.e. as $1/Q$. This leads to a trade-off between the comb pump-efficiency and stability, and its span.

In this Letter, we report a different configuration of a four-wave mixing comb using two pumps coupled to two WGMs one or several FSRs apart. Instead of coupling two vacuum and two (degenerate) pump fields, the leading-order interaction couples three pump fields and one vacuum field. For example, on the energy diagram in Fig.~\ref{fig:setup}(b) two pump photons with frequency $\omega_1$ are absorbed, and one pump photon with frequency $\omega_2>\omega_1$ is emitted, along with the photon with $\omega_-=2\omega_1-\omega_2$. A symmetric process will lead to generation of a photon with $\omega_+=2\omega_2-\omega_1$.
Since the pump field is much stronger than the vacuum field, the non-degenerate four-wave mixing is expected to be much more efficient than the degenerate one. Furthermore, as we will see below, the non-degenerate process is thresholdless. It occurs at any pump power, which allows one to avoid the undesirable high-power effects such as thermorefractive oscillations \cite{Matsko05} and stimulated Raman scattering. 

High efficiency and absence of a threshold lead to a cascade process in which the new frequency components generate further components, and a frequency comb arises. The spacing of this comb is equal to $\omega_2-\omega_1$, which can be a multiple of the FSR. The relative frequency drift of this comb components is determined by the drift of the lasers beat note, and is independent of $Q$, which resolves the stability - span dilemma. Phase-locking the lasers beat note to a high-stability reference oscillator one can achieve high comb stability even with low-Q resonators that are suitable for broader-range combs.

Our experimental setup diagram is shown in Fig.~\ref{fig:setup}(c). We used a fluorite resonator with $13.56$ GHz FSR. Fluorite has nearly linear dispersion at the wavelength of 1.5 $\mu$m, so the resonator spectrum around this wavelength is highly equidistant. Light was coupled in and out of the resonator using two optical fibers polished at the optimal coupling angle \cite{ilchenko99}.  The light from the input fiber reflected from the resonator's rim was collected by a photo detector to observe the spectrum of the resonator. 

The input fiber combined light from two lasers centered at 1560 nm. Both laser's frequencies were simultaneously scanned around the selected WGMs of the same family. This was achieved by fine-tuning each frequency until the selected resonances observed by the photo detector overlap. The resonator quality factor $Q=10^8$ (loaded) was made relatively low to increase the linewidth and to make the overlap easier to achieve. An optical spectrum analyzer (OSA) is connected to the output fiber for continuous data acquisition. The OSA was set to retain the peak power values, therefore a trace recorded for a sufficiently long period of time represented the situation when both lasers are fully coupled to the WGMs. 

\begin{figure}[htp]
\centerline{
\input epsf
\setlength{\epsfxsize}{3.5in} \epsffile{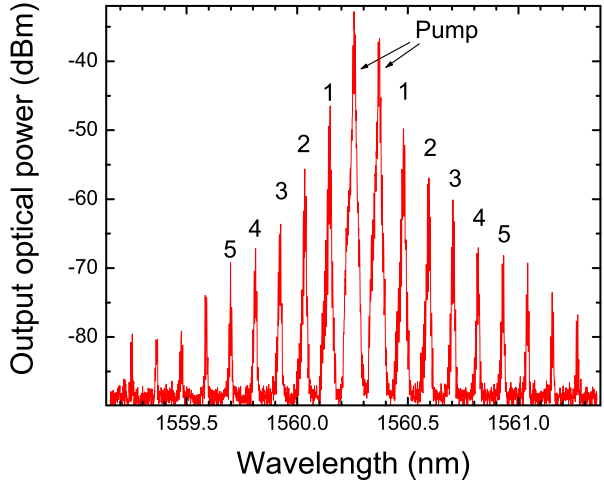} }
\vspace*{-0.2in}
\caption[]{\label{fig:1FSR} A typical spectrum of a bichromatically-pumped fluorite WGM resonator. The two highest peaks correspond to the two pumps, each of 2.9 mW at the input.}
\end{figure}

\begin{figure}[htp]
\centerline{
\input epsf
\setlength{\epsfxsize}{3.5in} \epsffile{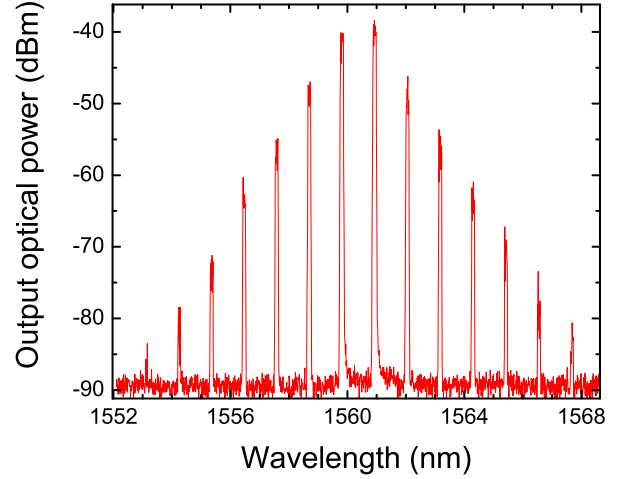} }
\vspace*{-0.2in}
\caption[]{\label{fig:10FSR} The spectrum for $P_1=P_2=4.75$ mW and 10 FSR (135.6 GHz) pump separation. }
\end{figure}

A typical comb generated by two pumps of $P_1=P_2=2.9$ mW set one FSR apart is shown in Fig.~\ref{fig:1FSR}. We did not see any pump power threshold required for the onset of the nonlinear oscillations. This measurement was repeated for the laser frequencies set two, three and ten FSRs apart. We had to stop at ten FSR because the lasers could not be tuned further. No significant difference between these measurements and the first one was observed. A comb spectrum corresponding to the ten FSR separation and the pump power $P_1=P_2=4.75$ mW is shown in Fig.~\ref{fig:10FSR}. The output pump power in Figs.~\ref{fig:1FSR} and \ref{fig:10FSR} is much less than the input powers because the output coupling was made weak so as not to overload the resonator. 

Our experiment can be described within the general theoretical model of Kerr-coupling of the WGMs \cite{Matsko05}. The difference in the present treatment is that the partial degeneracy of the Hamiltonian in \cite{Matsko05} is removed, and the leading-order process is a four-mode interaction rather than three-mode. The Hamiltonian $H=H_0+V$ describing our system includes the following terms:
\begin{eqnarray} \label{ham}
H_0& =& \hbar \omega_a a^\dag a+ \hbar \omega_b
b^\dag b + \hbar \omega_c c^\dag c+ \hbar \omega_d d^\dag d,\nonumber \\  
V &=&-\hbar \frac{g}{2} :( a+ b+c+d+h.c.)^4:,
\end{eqnarray}
where $c$, $a$, $b$, and
$d$ are photon annihilation operators at frequencies $\omega_c<\omega_a<\omega_b<\omega_d$,
respectively. The coupling constant $g$ is found in \cite{Matsko05}:
\begin{equation} \label{g}
g = \omega \frac{n_2}{n_0} \frac{\hbar \omega c}{{\cal V}
n_0},
\end{equation}
where $n_2$ is the Kerr nonlinearity of the resonator material, $n_0$ is its refraction index and ${\cal V}$ is the mode volume.

Notice that the fields represented by the operators $a$ and $b$ are strong pump fields, while the fields represented by $c$ and $d$ are initially vacuum fields. In Hamiltonian (\ref{ham}) we will only retain the leading terms with at least three operators representing strong fields. Following the steps of  \cite{Matsko05} and introducing the Langevin formalism, we arrive at the following stationary state equations:
\begin{eqnarray} \label{oeq1b}
\gamma a &=& 6ig( a^\dag aa+2ab^\dag b+2a^\dag bc+bbd^\dag)+ f_a,\nonumber\\ 
\label{oeq2b}  
\gamma b &=& 6ig(b^\dag bb+2ba^\dag a+2b^\dag ad+aac^\dag) +f_b,\nonumber\\
\label{oeq3b} 
\gamma c &=&  6ig aab^\dag+f_c,\\
\label{oeq4b} 
\gamma d &=& 6ig bba^\dag+f_d,\nonumber
\end{eqnarray}
where the loss rate $\gamma$ and optical frequency $\omega$ are assumed to be the same for all modes. The Langevin forces 
\begin{equation} \label{aver1a}
\langle f_{a,b} \rangle = \sqrt{\frac{2\gamma P_{a,b}}{\hbar
\omega}} \ , \quad \langle f_{c,d} \rangle =0
\end{equation}
represent the external pump coupled into the WGMs. We have assumed critical input coupling, so that the coupling rate $\gamma$ in (\ref{aver1a}) and the loss rate $\gamma$ in (\ref{oeq1b}) are equal. As we have mentioned, the output coupling rate $\gamma_{out}$ is much less than $\gamma$, so we have neglected the terms with $\gamma_{out}$ in Eqs.~(\ref{oeq1b}). The output coupling rate determines the relation between the average of the intracavity field operator, e.g. $\langle a \rangle$, and the output power supplied to the OSA: 
\begin{equation} \label{Pout}
P_a^{out} = \langle a \rangle^2\gamma_{out}\hbar
\omega/2.
\end{equation}

In the following we will assume equal pump powers: $P_a=P_b=P$.  The solutions to the first pair of Eqs.~(\ref{oeq1b}) retaining the leading-order self- and cross-phase modulation terms are
\begin{eqnarray} \label{averab}
\langle a \rangle& =&\langle b \rangle = \sqrt{\frac{\gamma}{6g}F(P/P_{0})},\\
\label{psat}
P_{0}&\equiv&\frac{\gamma^2\hbar\omega}{12g}
=\frac{2\pi{\cal V}}{3n_2\lambda}\left(\frac{n_0}{4Q}\right)^2\approx\frac{1}{18}P_{th},
\end{eqnarray}
where $F(P/P_{0})$ is the real root of the cubic equation $9F^3+F-P/P_{0}=0$. The scaling power $P_{0}$ is expressed via the single-pump hyperparametric oscillations threshold power $P_{th}$, introduced in \cite{Matsko05}. Assuming $\lambda=1.56\,\mu$m, $n_0= 1.44, \, n_2=3.2\cdot 10^{-16}$ cm$^2$/W, $Q\approx 10^{8},\, {\cal V}\approx 10^{-4}$ cm$^3$, we find for our resonator $P_{0}\approx 50$ mW. 

Remarkably, the low-power limit of the function $F(P/P_{0})$ is $P/P_{0}$, which is also the solution of the first pair of Eqs.~(\ref{oeq1b}) with all nonlinear terms neglected. At higher powers $F(P/P_{0})$ grows \emph{slower} than $P/P_{0}$, which is indicative of the ``pumping inefficiency" \cite{kippenberg04ieee} due to self- and cross- phase modulation of the pump fields.

From the last pair of Eqs.~(\ref{oeq1b}) we find, using Eqs.~(\ref{averab}) and (\ref{Pout}), the intracavity field operators mean values and the output powers for the new optical fields. Notice that the following solutions have no pump power threshold:
\begin{eqnarray} \label{cd}
\langle c\rangle &= &\langle d\rangle=\frac{6g}{\gamma}\langle a\rangle^3=\sqrt{\frac{\gamma}{6g}}F^{3/2}(P/P_{0}),\\
\label{Pcd}
P^{(1)}&=&P_{c}=P_{d}= \frac{\gamma_{out}\hbar\omega}{2}\frac{\gamma}{6g}F^{3}(P/P_{0}).
\end{eqnarray}

The output pump and signal powers delivered to the OSA are given by Eqs.~(\ref{Pout}) and (\ref{Pcd}), respectively. This allows us to determine $F(P/P_{0})$ at e.g. $P=0.75$ mW, and therefore to find the experimental value of $P_0$. We have carried out this measurement for one, two, three and ten FSR pump separation and found $P_0=10\pm 1$ mW. This value is a factor of five below our theoretical estimate. The discrepancy is likely due to large uncertainty in the shape of the hand-polished resonator, and the consequent order-of-magnitude uncertainty in the mode volume ${\cal V}$. Also, the intrinsic Q of the resonator could be larger that the measured loaded value.

\begin{figure}[htp]
\centerline{
\input epsf
\setlength{\epsfxsize}{3.3in} \epsffile{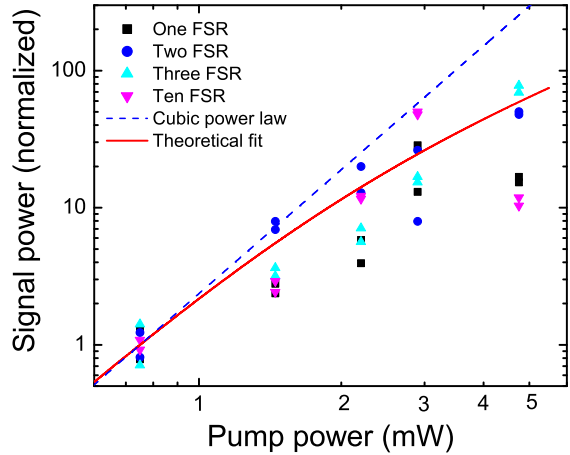} }
\vspace*{-0.3in}
\caption[]{\label{fig:SidebandP1} Normalized power of the first-order frequency components vs. the pump power: the data and the theory curves.
}
\end{figure}

To compare the theoretical prediction (\ref{Pcd}) with the experiment we have measured the ratio $P^{(1)}(P)/P^{(1)}(P=0.75\,{\rm mW})$ and plotted it as a function of the input pump power $P$, see Fig.~\ref{fig:SidebandP1}. Also shown are the theoretical curve (\ref{Pcd}) and the cubic power curve, which would be the solution in the absence of the ``pumping inefficiency". We see that the role of this inefficiency is significant.

\begin{figure}[htp]
\centerline{
\input epsf
\setlength{\epsfxsize}{3in} \epsffile{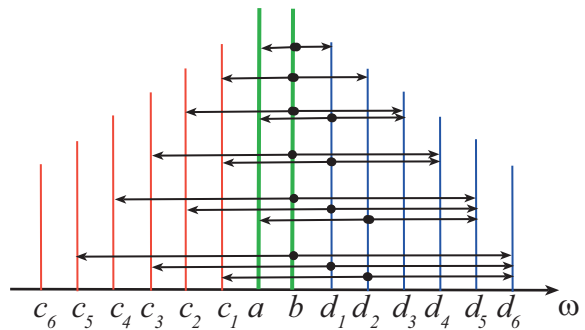} }
\caption[]{\label{fig:comb} The dominant channels for cascading the optical comb from a bichromatic pump. Every pair of arrows point from a mode in which a pair of photons is annihilated to the modes in which the photons are created.
}
\end{figure}

We have already mentioned that the bichromatically pumped process is thresholdless and therefore can be efficiently cascaded, with each excited mode playing the role of a new pump. Carrying out the above analysis recursively for the higher-order comb components, we find the following relation between their powers: 
\begin{equation} \label{Pn}
P^{(n)}/P^{(1)}=\xi_n\left( P^{(2)}/P^{(1)}\right)^{n-1}.
\end{equation}
The result (\ref{Pn}) describes the comb shape in the normalized form that can be easily verified in experiment. The factor $\xi_n$ has been introduced to account for multiple leading-order channels available for generation of sufficiently high-order frequency components, see Fig.~\ref{fig:comb}. Each channel shown in Fig.~\ref{fig:comb} annihilates a photon pair in some blue-shifted mode $m$ and couples it to a yet un-populated blue-shifted mode $n>m$ and to already populated red-shifted mode $n-2m-1<n$. A similar set of diagrams can be drawn for the red-shifted components.

The coefficients $\xi_n$ can be determined by calculating the powers delivered to a particular mode $n$ via all possible channels, and adding these powers up. Notice that by adding the optical powers, instead of fields, we disregard interference between different channels. This approximation may be suitable when the number of such channels is large. Requesting that the sum should be equal to $P^{(n)}$, we arrive at the following recursive relation: 
\begin{equation} \label{recursive}
\xi_n=\sum_{m=0}^{(n-1)/2}
\xi_m^{2}\xi_{n-2m-1},
\end{equation}
which yields a sequence $\xi_n$= 1, 1, 2, 3, 5, 8, 16, 27, ... 

\begin{figure}[htp]
\centerline{
\input epsf
\setlength{\epsfxsize}{3.5in} \epsffile{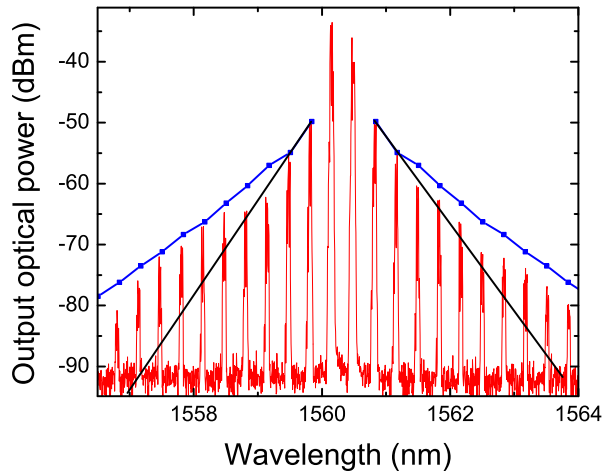} }
\vspace*{-0.2in}
\caption[]{\label{fig:combfit}A three-FSR-spaced comb generated by two 4.75 mW pumps is fit by the envelope functions based on a single-channel excitation (solid straight line) and on multiple excitations channels (line with dots.)   
}
\end{figure}

Equation (\ref{Pn}) allows us to fit the comb envelope, see Fig.~\ref{fig:combfit}. The comb shown in this figure is generated by two pumps of 4.75 mW each, set three FSRs apart. Also shown are the multi-channel envelope (\ref{Pn}) and the exponentially decaying envelope expected for single-channel excitation (when all $\xi_n=1$). We see that the multi-channel model approximates the observed comb much better than the single-channel one.

Let us point out that the processes described by Eq.~(\ref{oeq1b}) will also take place in the single-pump four-wave mixing process as soon as the threshold is exceeded and the comb arises. Moreover, given that $P_0\ll P_{th}$, processes (\ref{oeq1b}) will dominate the further comb dynamics. In particular, they may be expected to build up coherence between the comb components, which could not be expected from multiple \emph{independent} processes shown in diagram Fig.~\ref{fig:setup}(a) for various $N$. A high-purity RF beat note reported in \cite{savchenkov04,savchenkov08,del'haye07} indicates the presence of such a coherence.

To summarize, we have demonstrated a new and highly efficient method of generating optical combs with a relatively low-$Q$ WGM resonator and only a few milliwatt of the external CW pump power. The absence of a pump threshold in this process has been confirmed. The theoretical predictions for the shape and pump power dependence of the bichromatically-pumped comb reasonably agree with the experiment.  
We demonstrated the comb spacing varying from one to ten FSRs without any significant change in the comb behavior. This suggests that the spacing could be made even larger. Variable spacing of the comb may be convenient for the spectroscopy applications. We also would like to point out an interesting possibility of creating Moire comb pattern, by using three or more unequally spaced pump frequencies. This may enable an interesting approach to optical synthesizers.

This research was carried out by the Jet
Propulsion Laboratory, California Institute of Technology, under a
contract with the National Aeronautics and Space Administration. Dmitry Strekalov thanks Drs. Andrey Matsko and Anatoliy Savchenkov of OEwaves and Ivan Grudinin of Caltech for helpful discussions.

\end{document}